\documentclass[pra,twocolumn,showpacs]{revtex4}
\usepackage{graphics}

\draft
\begin{document}

\title{Formation of Vector Solitary waves with Mixed Dispersion  in Bose-Einstein Condensates.}
\author{L. Plaja, J. San Rom\'an}
 \affiliation{Departamento de F\'\i sica Aplicada, Universidad de
Salamanca,\\ E-37008 Salamanca, Spain}
\date{\today}

\begin{abstract}
We demonstrate the existence of a new class of two-component vector solitary waves in which dispersion coefficients have of opposite signs. Stability is achieved by inclusion of an additional linear coupling between the vector components that counterbalances the instability produced by the mixed dispersion and the non-linearity. In addition, we demonstrate that these solutions are experimentally observable as gap vector solitons in Bose-Einstein condensates located in oscillating optical lattices. The proposed experiment falls well within the standard experimental procedures for generating gap solitons in today's laboratories.
\end{abstract}

\pacs{03.75.Lm}
\maketitle

\section{Introduction}

Dispersion is a fundamental process in the dynamics of wavepackets. Its effect is revealed dramatically as the expansion of the wavepacket's size in time. For practical applications this extension becomes a serious limitation:  degrades communication bit rates, and prevents efficient methods for storage and transport of classical or quantum information. Fortunately, this situation changes radically in non-linear systems, where the non-linear potential can counteract dispersion and may result in a preservation of the wavepacket shape in time. The characterization of situations where these {\em soliton} solutions emerge constitutes a fundamental challenge for research. Besides its mathematical interest, especially relevant is the description of situations in which solitons or soliton-like structures appear in Nature, opening the door for technological applications (as happened with electromagnetic waves in Kerr media  \cite{molle80A}). 

The non-linear Schr\"odinger equation (NLSE) constitutes one paradigmatic case for the study of solitons \cite{zakha71A,haseg73A}. It describes different physical systems, among them the mean field dynamics of Bose-Einstein condensates (BEC). Bright solitons, or more generally solitary waves, in BEC have been demonstrated experimentally for the case of attractive interactions \cite{khayk02A,strec02A} and dark solitons in the case of repulsive interactions \cite{burge99A,densc00A}. In addition, bright solitons can be also obtained for repulsive interactions in condensates near the edge of the Brillouin zone of an optical lattice (gap solitons) . In this case, a condensate in the lowest band will experience a negative effective mass, thus inverting the sign of dispersion, which now can be counterbalanced by the repulsive non-linear potential. The first experimental observation of this type of solitary waves has been reported recently \cite{eierm04A}. 

Vector solitons may appear in situations where two or more different wave modes can coexist. In this case, the soliton is composed by a  set of wavepackets, one for each mode that propagate jointly. Every mode, therefore, contributes to the overall non-linear potential that allows for self-trapping.  Although the integrability of the resulting system of equations is not ensured for  the general two-component case, special cases as biased photorefractive crystals can be demonstrated to enclose Manakov solitons as solutions \cite{manak74A}. Other examples of two-component vector solitary waves can be found in waveguide arrays, where the two radiation modes correspond to different energy bands \cite{mande03A,mande03B}. In the general case, each equation of the set describing the vector soliton includes the counter-balance between dispersion and non-linearity. Therefore each vector component, even  evolving uncoupled, has the dynamical conditions to reach a soliton wave, which grants in great manner the stability of the coupled system. However, this condition is quite restrictive, since one may think in situations in which not all the vector components are described by equations in which the non-linearity and the dispersion balances. A particular case is the formation of vector gap solitons in which the geometry of the bands at the edge of the Brillouin zone only leads to a negative sign in the dispersion of the lowest band. In the uncoupled case, therefore, only the vector component corresponding to the lowest band will evolve to a soliton, while the other components (belonging to the upper bands) will rapidly disperse. In this case, the coupling is a fundamental aspect for the formation of a localized vector wavepacket. The aim of this letter is, first, to propose for the first time a set of two coupled NSE with opposite signs in the dispersion term which sustains localized self-trapped vector solutions,  and, second, to demonstrate that such solitary solutions can be observed in physical systems as BEC in oscillating optical lattices. 

To begin, let us consider the following system of coupled NLSE equations: 
\begin{eqnarray}
i { {\partial u_1} \over {\partial \tau}} & = & - { \alpha  \over 2}  \left [  {1 \over 2} \Delta k +i { {\partial }  \over {\partial \eta}}  \right ]^2 u_1 +  \beta u_2 +   \left( \gamma_{1} |u_1|^2 +  |u_2|^2 \right ) u_1  \nonumber \\
\label{eq:syssol1} \\
i { {\partial u_2} \over {\partial \tau}} & = & \delta u_2+ { \alpha^{-1}  \over 2}  \left [  {1 \over 2} \Delta k -i { {\partial }  \over {\partial \eta}}  \right ]^2 u_2+  \beta^* u_1 + \nonumber \\
& &   \left(  |u_1|^2 +\gamma_{2} |u_2|^2  \right ) u_2  \nonumber \\
\label{eq:syssol2}
\end{eqnarray}
Where $\alpha$, $\gamma_1$ and $\gamma_2$ are positive. $\alpha$ accounts for the mass difference between the  two vector components, $\Delta \kappa$ accounts for the difference between the group velocities of the vector components evolving freely (i.e. without coupling), and $\delta$ accounts for the gap between the energy dispersion curves of each component. These two later parameters are of no fundamental importance, as we have demonstrated selftrapping also for the case $ \Delta \kappa = \delta =0$, but should be taken into account to describe particular physical systems as BEC in optical lattices, for which each component belongs to a different energy band. In many aspects, Eqs. (\ref{eq:syssol1}-\ref{eq:syssol2}) resemble to those describing the formation of solitary waves in light passing through a periodically twisted birrefringent fiber \cite{wabni91A}. In both cases, the vector components interact through linear and non-linear couplings, and velocity and energy mismatch between the vector components are taken into account. However, in contrast with Eqs. (\ref{eq:syssol1}-\ref{eq:syssol2}), the dispersion terms in \cite{wabni91A}  have the same signs and the non-linear potentials are attractive, i. e. dispersion balances the non-linearity in each equation. Up to our knowledge, this characteristic is shared by the generality of models for vector solitons. Therefore, the novelty of the system (\ref{eq:syssol1}-\ref{eq:syssol2}) is the opposite sign in the dispersion terms for the two components. As discussed above, in this case  the coupling is fundamental for the stabilization of the second component of the vector solitary wave. Fig. \ref{fig:sol_model} shows the possibility of self-trapping  through the numerical integration of Eqs. (\ref{eq:syssol1}-\ref{eq:syssol2}).  The initial wavefunction is a Gaussian in one of the vector components and zero in the other, therefore a situation quite different from the two-component solitary wave in which it finally evolves. The parameters are taken as $\alpha=0.775$, $\delta=0.0114$, $\beta=-0.014$, $\gamma_1=1.43$,  $\gamma_2=0.56$ and the norm of the initial wavefunction is $0.69$, this particular values come from estimations from the Bose-Einstein experiment computed below. Figure  \ref{fig:sol_model}(a) and (b)  correspond to the time evolution of the first and second component of the vector function, respectively. In both cases a transient time is observed before the system reaches a stable configuration. During this transient, the population oscillates between both vector components while radiation is emitted at both sides of the wavefunction. Matter radiation is associated to the transient and may have a fundamental role as a dissipation process that allows the system to converge to the stable solution \cite{plaja04A,yulin03A}. The final population enclosed in the solitary wave depends on this transient, which is more pronounced the greater the difference between the initial wavepacket and the final solitary wave is.

Let us now focus to the experimental realization of this kind of vector solitary waves. To this end we will  first demonstrate that Eqs. (\ref{eq:syssol1}-\ref{eq:syssol2}) can describe the evolution of a Bose-Einstein condensate in an oscillating optical lattice. Afterwards, as a further proof,  we will compute directly the time evolution of the condensate, as given by Gross-Pitaevskii equation. These {\em ab initio} computations will demonstrate  that, for conditions close to the experiment \cite{eierm04A}, vector solitary waves with mixed dispersion appear spontaneously.

Following the philosophy of  \cite{eierm04A}, we will consider a Bose-Einstein condensate prepared  near the edge of the Brillouin zone of an optical lattice. The condensate is tightly confined in the transversal direction by a trapping potential. Under these circumstances, the condensate dynamics can be considered one dimensional, provided $(\mu/\hbar \omega_\perp-1)/2<<1$ \cite{plaja02A}, $\mu$ being the chemical potential and $\omega_\perp$ the transversal frequency of the trap. The dynamical equation is then
\begin{equation}
\label{eq:gp}
i \hbar {{\partial } \over {\partial t}} \Phi(z,t) =  \left ( H_0+H_{nl}  \right ) \Phi(z,t)
\end{equation}
where $H_0= \hat{p}^2/2m + U(z,t)$, $H_{nl}=g_{1D} \left | \xi(z,t) \right |^2$, with $g_{1D} =N \hbar a_s/m a_\perp$, $N$ being the number of atoms, $a_s$ the scattering length of the condensed element (in our case $^{87}Rb$) and $a_\perp$ the transversal size of the wavefunction (assumed to be in the lowest state of the radial harmonic potential). The potential $U(z,t)$ corresponds to a time-dependent axial lattice. In our case this potential includes an oscillatory motion of frequency $\omega$, and amplitude $\zeta_0$. Therefore $U(z,t)=U_0[z-\zeta(t)]$ with $\zeta(t)= \zeta_0 \sin(\omega t)$, and $U_0(\varsigma)=V_0 \cos^2(\pi  \varsigma/a_0)$ the stationary lattice potential. 

Let us now define a new wavefunction $\phi$ according to the transformation
\begin{equation}
\phi(z,t)= \exp[i {\hat{p} \over \hbar} \zeta(t)] \Phi(z,t)= \Phi(z+\zeta(t))
\end{equation}
Eq. (\ref{eq:gp}) may now be written as
\begin{eqnarray}
\label{eq:gp_trans}
i \hbar {\partial \over {\partial t}} \phi(z,t) & = & \left [ {{\hat{p}^2} \over {2m}} - \dot{\zeta}(t)  \hat{p} \right ]  \phi(z,t) +U_0(z) \phi(z,t)+ \nonumber \\
 &   & g_{1D}  |\phi(z,t)|^2 \phi(z,t)
\end{eqnarray}
in which the time dependence is removed from the potential term and translated to a linear coupling term. Note that this equation is formally identical to the NLSE of a 1D system in interaction with a semiclassical electromagnetic field in dipole approximation, the effective vector potential being $A(t)=(mc/q) \dot{\zeta}(t)$. This optical analogy is fruitful in understanding lattice oscillations as direct interband transitions, in the  same fashion as electromagnetic waves produce optical transitions of electrons in crystals. This type of coherent transfer between bands has been experimentally characterized in \cite{hecke02A}. 

Now we express the wavefunction $ \phi(z,t)$ as a superposition of states of the first and second band. Using the Wannier basis \cite{ziman},
\begin{equation}
\label{eq:wannier_expansion}
\phi(z,t)= \sum_\ell e^{i k_0 \ell} \left [ f_1(\ell,t) a_1(z-\ell)+ f_2(\ell,t) a_2(z-\ell) \right ]
\end{equation}
where $k_0=m v_0/\hbar$ corresponds to the initial momentum of the wavefunction, located near the edge of the Brillouin zone. We may now follow the standard procedure for the effective mass approach in the form used, for instance, in \cite{plaja04A}. Substituting (\ref{eq:wannier_expansion}) in (\ref{eq:gp_trans}), and assuming negligible overlap between Wannier functions of neighbor sites (tight-binding approx.), we find the dynamical equations for the wavefunction envelopes
\begin{eqnarray}
\label{eq:ev_fp}
i \hbar & {\partial \over \partial t} & f_p(\xi, t) = \hat{T}_p f_p(\xi,t)-  \nonumber \\
 & & {{2 a_0}  \over \hbar} \dot{\zeta}  \sum_{q \neq p} \pi_{p,q} \left ( -i \hbar {\partial \over \partial \xi }  \right ) f_q(\xi, t)+  \nonumber \\
 & & g_{1D} \sum_{q, n,m} \Gamma_{p,q,n,m} f^*_q(\xi, t)  f_n(\xi, t)  f_m(\xi, t) 
\end{eqnarray}
with band labels $\{p,q,n,m\}$ valued $1$ or $2$, $\xi$ defined as  the continuous limit of $\ell$, and with $\pi_{p,q} =i \int dz a_p^*(z-\ell) (-i \hbar) \partial a_q(z-\ell-a_0)/\partial z$  and $\Gamma_{p,q,n,m} \equiv \int dz a_p^*(z-\ell) a_q^*(z-\ell) a_n(z-\ell) a_m(z-\ell) $ constant parameters. The kinetic operator is defined as
\begin{equation}
\hat{T}_p  =  \epsilon_p(k_0) + \left ( v_{g,p} - {{2 a_0 \pi_{p,p}}  \over \hbar} \dot{\zeta}  \right ) \left ( -i \hbar {\partial \over \partial \xi }  \right )- {\hbar^2 \over {2 m_p^*}} {\partial^2 \over {\partial \xi^2}} 
\end{equation}
with $\epsilon_p(k_0)$ the energy of the $p$ band evaluated at the point $k_0$ of the Brillouin zone and  $v_{g,p} \equiv (1/\hbar) (\partial/\partial k) \epsilon_p |_{k_0}$ and $1/m_p^*=(1/\hbar^2) (\partial^2/\partial^2 k) \epsilon_p |_{k_0}$  the group velocity and the effective mass. The linear coupling in Eq. (\ref{eq:ev_fp}) describes the direct (optical) interband transitions induced by the oscillating lattice. Its derivation follows from the evaluation of the linear coupling in Eq. (\ref{eq:gp_trans}) projected on the Wannier function $\exp{( ik_0 \ell)} a_p(z-\ell)$ which leads to sum of the type $\sum_{n, \ell'} \exp \left[i k_0 (\ell'-\ell)\right] f_n(\ell',t) \int dz a^*_p(z-\ell) \hat{p} a_n(z-\ell')$. Assuming that the leading term in the sum corresponds to the coupling between nearest neighbors, and since $k_0 a_0 \simeq \pi$, we can approximate this sum to $ -i \sum_n \pi_{p,n} \left[  f_n(\ell+a_0,t) -  f_n(\ell-a_0,t)  \right] $ which, in the continuous limit, can be written as $ -i 2 a_0 \sum_n \pi_{p,n} {\partial \over \partial \xi} f_n(\xi,t) $.

 We shall assume that the lattice oscillation is tuned near the resonance of the interband transition: $\hbar \omega = \epsilon_2(k_0) - m^*_2 v_{g,2}^2/2 -   \epsilon_1(k_0) + m^*_1 v_{g,1}^2/2 -\hbar \Delta$, $\Delta$ being a small quantity. Following the standard approach to resonance \cite{allen87A}, we define new rotating system wavefunctions $g_p$ according to 
\begin{eqnarray} 
\label{eq:def_g1}
f_1(\xi,t) & \equiv & \exp \left [ -i \left( \epsilon_1-{ {m_1^* v_{g,1}^2} \over 2} \right ) {t \over  \hbar} \right ]  \exp \left ( -i  \overline{k}  \xi  \right ) g_1(\xi,t) \nonumber \\
\\
f_2(\xi,t) & \equiv & \exp \left [ -i \left( \hbar \Delta + \epsilon_2-{ {m_2^* v_{g,2}^2} \over 2} \right ) {t \over  \hbar} \right ]  \times \nonumber \\
& & \exp \left ( -i \overline{k} \xi  \right ) g_2(\xi,t) 
\label{eq:def_g2}
\end{eqnarray}
with $\hbar \overline{k}= (m_1^* v_{g,1}+m_2^* v_{g,2})/2$. In  the philosophy of the rotating-wave approximation \cite{allen87A}, we discard oscillating terms after substituting  (\ref{eq:def_g1}-\ref{eq:def_g2}) into (\ref{eq:ev_fp}), thus  retaining only the slowly varying dynamics. Under these approximations, Eq. (\ref{eq:ev_fp}) can be finally cast to the form of Eqs . (\ref{eq:syssol1} - \ref{eq:syssol2}) defining adimensional variables $\eta=\xi/ a_0$ and $\tau=t/\tau_0$, with $\tau_0=\sqrt{|m_1^* m_2^*|} a_0^2/\hbar$, and defining the parameters as  
$\alpha=\sqrt{|m_2^*/m_1^*|}$, $\Delta k =a_0 (m_2^* v_{g,2}-m_1^* v_{g,1})/ \hbar$, $\delta=\Delta \tau_0$, $\beta=-a_0 \tau_0 A_0 \omega \pi_{1,2} \overline{k}/\hbar$, and $u_j=\sqrt{2g_{1D}\Gamma_{1,2} \tau_0/\hbar} g_j$ , and $\gamma_1=\Gamma_{1,1,1,1}/\Gamma_{1,1,2,2}$ and $\gamma_2=\Gamma_{2,2,2,2}/\Gamma_{1,1,2,2}$.  

To check the validity of our approach, we have also performed {\em ab initio} integrations of the evolution of a Bose-Einstein condensate in  an optical lattice, as given by Eq. (\ref{eq:gp}). In these computations, the transversal trapping is assumed to be constant in time and harmonic with $\omega_\perp=2 \pi \times 50 Hz$. Initially a condensate of 400 atoms is prepared in a trap formed by this transversal potential and an axial potential with $\omega_{||}=2 \times 9 Hz$, (point (a) in figure \ref{fig:esquemabandas}).  At the starting time, the axial potential is switched off, and the condensate is phase imprinted with a momentum $\hbar k_0=\hbar 0.9  \pi/a_0$ with $a_0=1.96 \mu m$. In the experiment \cite{eierm04A} this phase imprinting is replaced by a displacement of the optical lattice with constant velocity. After the phase imprinting,  an oscillating optical lattice potential is switched on adiabatically during $10 ms$ to reach a maximum potential of $V_0=0.85 E_r$ ($E_r=\hbar^2\pi^2/2ma_0^2$). As the lattice period is $a_0$, the wavepacket imprinted with momentum $\hbar k_0$  is now located near the edge of the first Brillouin zone (points (b) and (c) in figure \ref{fig:esquemabandas}). The frequency of the lattice oscillation is $\omega=2 \pi \times 64 Hz$ and the oscillation amplitude $\zeta_0=0.35 \mu m$. With this procedure we obtain a wavepacket with similar characteristics that in the experiment \cite{eierm04A} but in an oscillating lattice. Figs \ref{fig:numerical}(a)  and (b) show the results of this {\em ab initio} integration of Eq. (\ref{eq:gp}) in a similar fashion as in plots \ref{fig:sol_model}(a) and (b) respectively. In Fig. \ref{fig:numerical}, the vector components $f_1$ and $f_2$ have been extracted from the time-dependent wavefunction by projecting in the Wannier base of the first and second bands. The mechanism of formation of the solitary wave follows the same steps as in the model eqs. (\ref{eq:syssol1}-\ref{eq:syssol2}): a transient characterized by the emission of radiation, and final state in which the two-component solitary wave is formed. Note that, in contrast with Fig. \ref{fig:sol_model}, no oscillations between the vector components are observed. This is not a general case, as our computations show that the oscillating transient behavior can be recovered when the form of the initial wavefunction is chosen to be more similar to the solitary wave shape. We have found, in this case, that the frequency of the oscillations increases with the amplitude $\zeta_0$ and as frequency $\omega$ decreases (i.e. $\delta$ increases). This suggests that these oscillations are of the Rabi type. The absence of them in Fig.  \ref{fig:numerical} can be then explained by the more complex dynamics involved in the convergence of the system as the difference between the shape of the initial wavefunction and the final solitary wave is larger. For the case of scalar gap solitons, it is known that the wavepackets drifts in momentum space as a consequence of the radiation process during the transient \cite{plaja04A}. Under these circumstances, the parameters of the interband transitions change and, therefore, the Rabi regime is severely distorted.

In conclusion, we have proved the existence of a new kind of vector solitary wave in which the dispersion sign is opposite for each component. Under these circumstances, one of the vector elements is highly dispersive. Selftrapping  is obtained through a linear coupling in the vector equations, which allows the highly dispersive component to be stabilized. We have demonstrated that this kind of vector solitary waves can be found in experimentally using Bose-Einstein condensates in oscillating optical lattices.

This work has been supported by the Spanish Ministerio de Ciencia y Tecnolog\' \i a (FEDER funds, grant BFM2002-00033).

\newpage

\begin{figure}
\caption{Time evolution of a two-component wavepacket as described by the coupled system (\ref{eq:syssol1}-\ref{eq:syssol2}). The wavepacket is initially a Gaussian in the lowest component (a) and empty in the upper component (b). After a transient time in which population oscillates between both vector components and radiation takes place, a two-component vector solitary wave is reached. The vertical scale in (b) has been magnified a factor two form that of (a).}
\label{fig:sol_model}
\end{figure}

\begin{figure}
\caption{Band structure of the optical lattice considered in the text. The condensate is initially located at momentum 0 (a). After imprinting a momentum $\hbar k_0$ and the adiabatic switch on of the oscillating lattice, the wavefunction is found located at the edge of the Brillouin zone with components in the two lowest bands (b) and (c).}
\label{fig:esquemabandas}
\end{figure}

\begin{figure}
\caption{Time evolution of the condensate's wavepacket as computed from the Gross-Pitaevskii equation (\ref{eq:gp}). The wavepacket at $t=0$ corresponds to the lowest eigenstate of this equation  for  the trap defined in the text. At $t=0$ the condensate is imprinted with a momentum $\hbar k_0$. Subsequently an oscillating optical lattice is grown adiabatically. Plot (a) and (b) show the projection of the wavefunction onto the Wannier basis of the first and second band respectively. The vertical scale in (b) has been magnified a factor two form that of (a)}
\label{fig:numerical}
\end{figure}

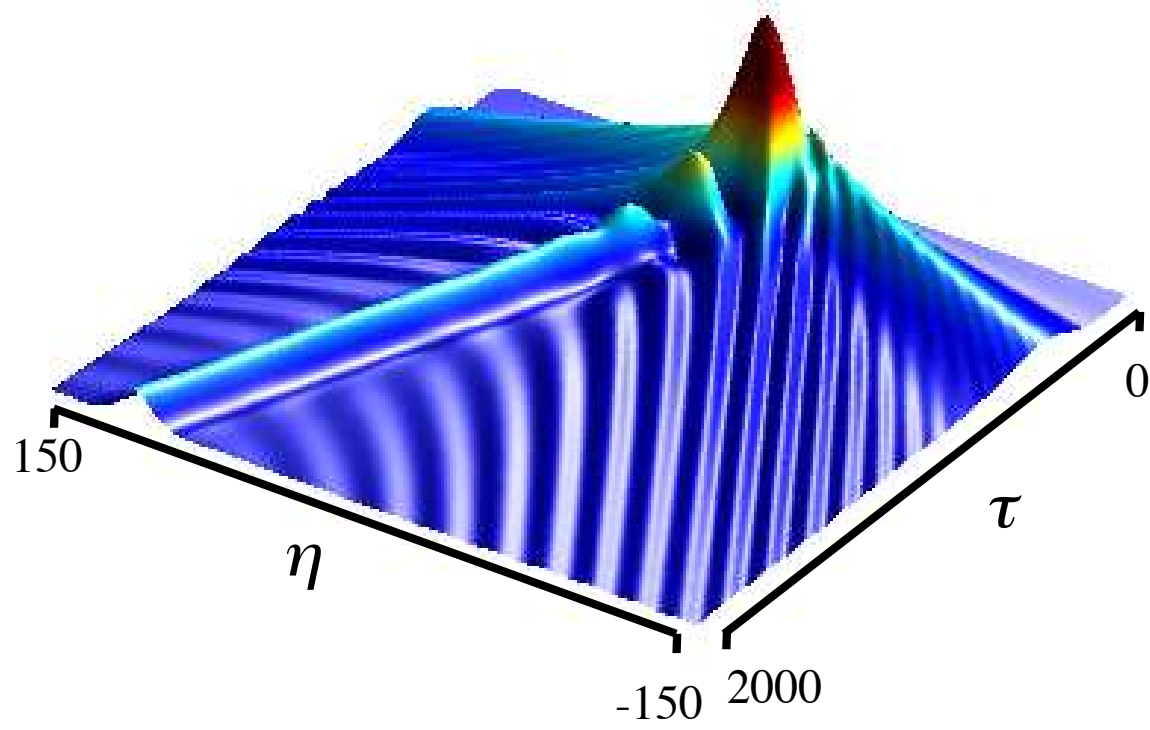
\end{document}